# ABRAHAM-BASED MOMENTUM AND SPIN OF OPTICAL FIELDS UNDER CONDITIONS OF TOTAL REFLECTION

Aleksandr Ya. Bekshaev

Odessa I.I. Mechnikov National University
Dvorianska 2, 65082 Odessa, Ukraine
E-mail: bekshaev@onu.edu.ua

This memo contains a collection of formulas describing the electromagnetic energy, momentum and spin distribution of an optical field formed in dispersion-free dielectric media separated by a plane interface when an incident monochromatic plane wave is totally reflected. The formulas are based on the Abraham momentum definition and include the momentum decomposition into the orbital (canonical) and spin parts as well as explicit dual-symmetric separation of the electric and magnetic contributions.

This material was prepared in February 2013 and reflects the situation as it could be perceived at that time. However, it had not been finalized and published because of the difficulties in physical interpretation of singular terms in the spin and orbital momentum expressions associated with the sharp interface. Meanwhile, it has become clear that the "naïve" Abraham approach is not correct for this problem and the electromagnetic spin and momentum in inhomogeneous media are better characterized by the more elaborated relations based on the Minkowski paradigm [see, e.g., Phys. Rev. A **83**, 013823 (2011); **86**, 055802 (2012); arXiv:1706.05493]. In application to the total-reflection situation this Minkowski-based description was recently illustrated in arXiv:1706.06263, so the present material is mainly of historical interest. However, it seems useful to make it known for a wide audience, at least for comparison with the recent improved approaches and for suitable references.

The last section, treating the ponderomotive action experienced by a Mie particle in the evanescent wave, is independent of the Abraham – Minkowski controversy. The numerical calculations preserve their validity, and association of the various force and torque components with corresponding components of the optical momentum and spin remains legal in the Minkowski pattern. The results of the last section were partly used in other published works [e.g., Nature Commun. **5**, 3300 (2014)].

## 1. Introduction

It is well known that a propagating plane wave carries momentum along its propagation direction (**k**-vector) [1]. There are many manifestations of this feature and the most immediate is perhaps the mechanical action: an electromagnetic wave pushes small absorbing particles in this direction due to the radiation force (see, e.g., reviews [2,3] and references therein). This effect is mostly independent of the wave polarization. On the other hand, an elliptically-polarized propagating plane wave also carries spin angular momentum along its **k**-vector and can spin a small absorbing particle about its axis. This effect is proportional to the polarization ellipticity and is also well studied [4–7].

The situation becomes much more intriguing if we consider an *evanescent* electromagnetic wave. First, many years ago it was noticed that a circularly polarized evanescent field contains an *ellipticity-dependent* energy flow (Poynting vector component) in the *transverse* direction [8–11]. Originally, it was assumed that this effect produces a transverse shift of the circularly-polarized beam totally-reflected from a dielectric interface. However, further results proved that this shift (also known as the spin Hall effect of light) is unrelated to evanescent waves and the Fedorov-Imbert transverse energy flow [12–17]. Thus, this anomalous transverse momentum has never been further studied in detail nor was it observed experimentally.

Second, recent considerations of energy flows and angular momenta in surface plasmon-polaritons at a metal-dielectric interface [18] reveal that any *linearly*-polarized evanescent electromagnetic wave carries spin angular momentum in the *transverse* direction. This is very unusual for electromagnetism and quantum mechanics of photons, which carry spin only *along* the **k**-vector. Note that such transverse spin arises from an imaginary longitudinal field component and the amplitude inhomogeneity of an evanescent wave. To the best of our knowledge, this extraordinary transverse spin has never been discussed and detected experimentally.

Thus, evanescent electromagnetic waves are expected to exhibit quite curious mechanical properties, which are in sharp contrast to the properties of usual propagating waves. Additional interest concerns the field momentum subdivision into the spin and orbital parts [19–21] in presence of the sharp boundary, emergence of the singular boundary spin and orbital energy flows and their interplay in the total energy transfer pattern [18]. All these facts need experimental verification and a natural way of their study is to exploit the well established mechanical action of momentum or angular momentum of light which manifests itself by the radiation force and torque that causes linear motion and spinning of a small probe particle immersed in the optical field. The possibilities and efficiency of this approach for the evanescent field were repeatedly confirmed in many related works (see, e.g., Refs. [22–28]); however, the detailed picture, particularly with account for the physical nature of the ponderomotive agent (field energy gradient, spin or orbital momenta of light) and for the special properties of the particle, is still the subject of discussion [29–33].

In this paper, we analyze the evanescent field structure, energy flow pattern and their possible mechanical action as it could be observed in experiment. In contrast to [18], we consider more practical situation of evanescent field formed in the low-density medium upon the total reflection phenomenon. The analysis is based on the recently developed complex-angle approach for calculation of optical forces and torques exerted on Mie particles in evanescent field [34] and is thus limited by situations where influence of the multiple scattering of light by the interface is negligible. This is admissible in many cases involving subwavelength particles, especially when the force or torque of interest acts parallel to the interface [25–27]; in more severe conditions, this approach gives good qualitative results and serves a base for a more thorough analysis.

## 2. Field configuration and the energy flow parameters

Consider a general model of evanescent field formation in the total-reflection of a monochromatic light wave with frequency $\omega$ where the electric and magnetic fields behave in time as $\mathrm{Re}\left[\mathbf{E}\exp\left(-i\omega t\right)\right]$, $\mathrm{Re}\left[\mathbf{H}\exp\left(-i\omega t\right)\right]$. Let the incident plane wave comes from the half-space $z < 0$ (medium 1) with parameters $\varepsilon_1$, $\mu_1$ and $n_1 = \sqrt{\varepsilon_1 \mu_1}$, and the boundary $z = 0$ separates it from the medium 2 ($z > 0$) with parameters $\varepsilon$, $\mu$ and $n = \sqrt{\varepsilon \mu}$. The wave direction is specified by the wave vector $\mathbf{k} = k_1\left(\cos\theta_1, 0, \sin\theta_1\right)^T$, $\theta_1$ is the angle of incidence, $k_1 = n_1\left(\omega/c\right)$ is the wavenumber in the medium 1, $c$ is the light velocity in vacuum.

### 2.1. Electromagnetic field components

1) The electric and magnetic vectors in the incident wave are described by equations

$$\mathbf{E}_0 = \begin{pmatrix} E_{0\parallel}\cos\theta_1 \\ E_{0\perp} \\ -E_{0\parallel}\sin\theta_1 \end{pmatrix} e^{ik_1(z\cos\theta_1 + x\sin\theta_1)}, \quad \mathbf{H}_0 = \sqrt{\frac{\varepsilon_1}{\mu_1}} \begin{pmatrix} -E_{0\perp}\cos\theta_1 \\ E_{0\parallel} \\ E_{0\perp}\sin\theta_1 \end{pmatrix} e^{ik_1(z\cos\theta_1 + x\sin\theta_1)} \quad (1)$$

where $E_{0\parallel}$ and $E_{0\perp}$ are the electric strength components parallel ($p$) and orthogonal ($s$) to the plane of incidence ($p$- and $s$-polarized components).

2) Reflected wave differs by replacement $\theta_1 \to \pi - \theta_1$ and by the transformed field amplitudes:

$$\mathbf{E}_r = \begin{pmatrix} -E_{r\|}\cos\theta_1 \\ E_{r\perp} \\ -E_{r\|}\sin\theta_1 \end{pmatrix} e^{ik_1(-z\cos\theta_1 + x\sin\theta_1)}, \quad \mathbf{H}_r = \begin{pmatrix} E_{r\perp}\cos\theta_1 \\ E_{r\|} \\ E_{r\perp}\sin\theta_1 \end{pmatrix} e^{ik_1(-z\cos\theta_1 + x\sin\theta_1)} \tag{2}$$

where

$$E_{r\perp} = R_\perp E_{0\perp}, \quad E_{r\|} = R_\| E_{0\|}, \tag{3}$$

and

$$R_\perp = \frac{\sqrt{\varepsilon_1/\mu_1}\cos\theta_1 - i\sqrt{\varepsilon/\mu}\sinh\alpha}{\sqrt{\varepsilon_1/\mu_1}\cos\theta_1 + i\sqrt{\varepsilon/\mu}\sinh\alpha}, \quad R_\| = \frac{\sqrt{\varepsilon/\mu}\cos\theta_1 - i\sqrt{\varepsilon_1/\mu_1}\sinh\alpha}{\sqrt{\varepsilon/\mu}\cos\theta_1 + i\sqrt{\varepsilon_1/\mu_1}\sinh\alpha} \tag{4}$$

are the reflection coefficients, $|R_\perp| = |R_\|| = 1$. Eqs. (4) follow from formal application of the well known Fresnel formulas and the Snell's law [1] to the total reflection case:

$$\sin\theta = \frac{n_1}{n}\sin\theta_1 = \cosh\alpha > 1, \quad \cos\theta = i\sqrt{\left(\frac{n_1}{n}\right)^2 \sin^2\theta_1 - 1} = i\sinh\alpha, \tag{5}$$

which corresponds to the complex refraction angle $\theta = \dfrac{\pi}{2} - i\alpha$.

3) Similarly, the refracted (evanescent) wave is described by relations [34]

$$\mathbf{E}_t = \begin{pmatrix} iE_{t\|}\sinh\alpha \\ E_{t\perp} \\ -E_{t\|}\cosh\alpha \end{pmatrix} \exp\left(ikx\cosh\alpha - kz\sinh\alpha\right),$$

$$\mathbf{H}_t = \sqrt{\frac{\varepsilon}{\mu}} \begin{pmatrix} -iE_{t\perp}\sinh\alpha \\ E_{t\|} \\ E_{t\perp}\cosh\alpha \end{pmatrix} \exp\left(ikx\cosh\alpha - kz\sinh\alpha\right) \tag{6}$$

where $k = n(\omega/c)$ is the wavenumber in the medium 2,

$$E_{t\perp} = T_\perp E_{0\perp}, \quad E_{t\|} = T_\| E_{0\|}, \tag{7}$$

and

$$T_\perp = \frac{2\sqrt{\varepsilon_1/\mu_1}\cos\theta_1}{\sqrt{\varepsilon_1/\mu_1}\cos\theta_1 + i\sqrt{\varepsilon/\mu}\sinh\alpha}, \quad T_\| = \frac{2\sqrt{\varepsilon_1/\mu_1}\cos\theta_1}{\sqrt{\varepsilon/\mu}\cos\theta_1 + i\sqrt{\varepsilon_1/\mu_1}\sinh\alpha}. \tag{8}$$

Noteworthy, the field of Eqs. (6) can be considered as a plane wave with wavevector $\mathbf{k}$ that lies in the ($XZ$) plane but possesses complex $z$-component [18,34]

$$\mathbf{k} = \left(k\cosh\alpha, 0, ik\sinh\alpha\right). \tag{9}$$

## 2.2. Field energy, momentum and angular momentum

Now consider parameters of the energy and momentum distribution in both regions 1 and 2. Generally, the electromagnetic momentum density and the energy density in the medium with parameters $\varepsilon_m$ and $\mu_m$ are determined by the known relations [1]

$$\mathbf{p} = \frac{g}{c}\mathrm{Re}\left(\mathbf{E}^* \times \mathbf{H}\right), \tag{10}$$

$$w = \frac{g}{2}\left(\varepsilon_m |\mathbf{E}|^2 + \mu_m |\mathbf{H}|^2\right) \tag{11}$$

where $g = (8\pi)^{-1}$ in the Gaussian system of units. In turn, the field momentum can be decomposed into the spin and orbital parts [19–21],

$$\mathbf{p}_S = \frac{g}{4\omega} \operatorname{Im} \left[ \nabla \times \left( \frac{1}{\mu_m} \mathbf{E}^* \times \mathbf{E} \right) + \nabla \times \left( \frac{1}{\varepsilon_m} \mathbf{H}^* \times \mathbf{H} \right) \right], \tag{12}$$

$$\mathbf{p}_O = \mathbf{p} - \mathbf{p}_S = \frac{g}{2\omega} \operatorname{Im} \left[ \frac{1}{\mu_m} \mathbf{E}^* \cdot (\nabla) \mathbf{E} + \frac{1}{\varepsilon_m} \mathbf{H}^* \cdot (\nabla) \mathbf{H} \right.$$

$$\left. + \frac{1}{2\mu_m^2} \nabla \mu_m \times \left( \mathbf{E}^* \times \mathbf{E} \right) + \frac{1}{2\varepsilon_m^2} \nabla \varepsilon_m \times \left( \mathbf{H}^* \times \mathbf{H} \right) - \frac{1}{\varepsilon_m \mu_m} \mathbf{E} \left( \mathbf{E}^* \cdot \nabla \varepsilon_m \right) - \frac{1}{\varepsilon_m \mu_m} \mathbf{H} \left( \mathbf{H}^* \cdot \nabla \mu_m \right) \right]. \tag{13}$$

The spin momentum density $\mathbf{p}_S$ (12) is related with the spin angular momentum density

$$\mathbf{S} = \frac{g}{2\omega} \operatorname{Im} \left( \frac{1}{\mu_m} \mathbf{E}^* \times \mathbf{E} + \frac{1}{\varepsilon_m} \mathbf{H}^* \times \mathbf{H} \right) \tag{14}$$

via equation [18,35]

$$\mathbf{p}_S = \frac{1}{2} \nabla \times \mathbf{S} . \tag{15}$$

Note that this form of equations allows for the possible spatial variability of the medium parameters $\mu_m$ and $\varepsilon_m$. The third and fourth terms in the brackets of Eq. (13) describe the special kind of the orbital momentum that, in contrast to the common notion [19,20], emerges due to elliptic polarization of the optical field provided that the medium is not homogeneous.

We start the calculation with the energy flow density in regions 1 and 2 via Eq. (10). In the medium 1 ($-\infty < z < 0$), where the field is formed by superposition of two plane waves (1) and (2), we have to substitute $\mathbf{E} = \mathbf{E}_0 + \mathbf{E}_r$, $\mathbf{H} = \mathbf{H}_0 + \mathbf{H}_r$, $\varepsilon_m = \varepsilon_1$, $\mu_m = \mu_1$; in the medium 2, the quantities to be substituted are $\mathbf{E} = \mathbf{E}_t$, $\mathbf{H} = \mathbf{H}_t$, $\varepsilon_m = \varepsilon$, $\mu_m = \mu$, see Eqs. (6). As a result, we obtain for the components of the total momentum density (10)

$$p_x^+ = \frac{gn}{c\mu} \cosh \alpha \left( |T_\perp E_{0\perp}|^2 + |T_\parallel E_{0\parallel}|^2 \right) e^{-2kz \sinh \alpha} \quad (0 < z < \infty); \tag{16}$$

$$p_y^+ = 2 \frac{gn}{c\mu} \sinh \alpha \cosh \alpha \operatorname{Im} \left( T_\parallel T_\perp^* E_{0\parallel} E_{0\perp}^* \right) e^{-2kz \sinh \alpha} \quad (0 < z < \infty); \tag{17}$$

$$p_x^- = 2 \frac{gn_1}{c\mu_1} \sin \theta_1 \left[ |E_{0\parallel}|^2 + |E_{0\perp}|^2 + \operatorname{Re} \left( |E_{0\perp}|^2 R_\perp^* e^{2ik_1 z \cos \theta_1} + |E_{0\parallel}|^2 R_\parallel^* e^{2ik_1 z \cos \theta_1} \right) \right] \quad (-\infty < z < 0); \tag{18}$$

$$p_y^- = -\frac{gn_1}{c\mu_1} \sin 2\theta_1 \operatorname{Re} \left[ E_{0\parallel} E_{0\perp}^* \left( R_\perp^* e^{2ik_1 z \cos \theta_1} - R_\parallel^* e^{-2ik_1 z \cos \theta_1} \right) \right] \quad (-\infty < z < 0); \tag{19}$$

$$p_z = 0 \quad (-\infty < z < \infty) \tag{20}$$

(superscripts '+' and '−' denote the positive $z > 0$ and negative $z < 0$ half-spaces). These results are well known (see, e.g., Ref. [11] where they were used for explanation of the Goos–Hänchen and Imbert–Fedorov beam shifts); however, the energy flow pattern they describe was considered in detail only recently [18]. Now we generalize that analysis to the total-reflection situation.

Importantly, according to Eqs. (16) – (20), only the "trivial" normal to the boundary component $p_z$ of the momentum, which is identical zero, is continuous in the whole space. Functions $p_x(z)$ (16), (18) and $p_y(z)$ (17), (19), characterizing the energy flow density, experience simple step discontinuities at $z = 0$. By using Eq. (5) and easily derived interrelations between the reflection (4) and transmission (8) coefficients,

$$\operatorname{Re}\left(1 + R_\perp\right) = \frac{1}{2} |T_\perp|^2, \quad \operatorname{Re}\left(1 + R_\parallel\right) = \frac{1}{2} \frac{\varepsilon \mu_1}{\varepsilon_1 \mu} |T_\parallel|^2, \quad R_\perp^* - R_\parallel = \frac{i}{2} \frac{\sinh \alpha}{\cos \theta_1} \left( 1 + \frac{\varepsilon \mu_1}{\varepsilon_1 \mu} \right) T_\parallel T_\perp^*,$$

these steps can be represented in a compact physically transparent form:

$$\left(p_y^+ - p_y^-\right)_{z=0} = -\frac{g}{c}\frac{n}{\mu}\sinh\alpha\cosh\alpha\,\mathrm{Im}\left(T_\perp^* T_\| E_{0\perp}^* E_{0\|}\right)\left(\frac{\mu}{\mu_1}+\frac{\varepsilon}{\varepsilon_1}-2\right), \tag{21}$$

$$\left(p_x^+ - p_x^-\right)_{z=0} = \frac{g}{c}\frac{n}{\mu}\cosh\alpha\left[\left|T_\perp E_{0\perp}\right|^2\left(1-\frac{\mu}{\mu_1}\right)+\left|T_\| E_{0\|}\right|^2\left(1-\frac{\varepsilon}{\varepsilon_1}\right)\right]. \tag{22}$$

Before proceeding further, note that, contrary to the total field momentum (10), where the electric and magnetic fields are entangled, the field energy (11), the spin (12) and orbital (13) momenta as well as the spin density (14) can be considered as sums of distinctly separable electric and magnetic contributions expressed by the first and second summands of corresponding expressions. This rule applies also to the energy flow steps (21) and (22) where summands proportional to $1-\mu/\mu_1$ ('magnetic') and those proportional to $1-\varepsilon/\varepsilon_1$ ('electric') can be easily singled out. This division is not formal and reflects specific features of the electromagnetic field interaction with material objects which usually possess preferential sensitivity to electric or magnetic field [24,30,33]. For this reason, we will highlight these partial contributions by superscripts 'e' and 'm'.

Our main interest is associated with the spin density (14). For the evanescent field, in the region $0 < z < \infty$ Eqs. (6) and (14) give

$$S_x^{e+} = S_x^{m+} = -\frac{g}{\omega\mu}\cosh\alpha\,\mathrm{Im}\left(T_\| T_\perp^* E_{0\|} E_{0\perp}^*\right)e^{-2kz\sinh\alpha}, \quad S_x^+ = 2S_x^{e+} = 2S_x^{m+}; \tag{23}$$

$$\left\{\begin{matrix} S_y^{e+} \\ S_y^{m+} \end{matrix}\right\} = -\frac{g}{\omega\mu}\sinh\alpha\cosh\alpha\left\{\begin{matrix} \left|T_\| E_{0\|}\right|^2 \\ \left|T_\perp E_{0\perp}\right|^2 \end{matrix}\right\}e^{-2kz\sinh\alpha}, \quad S_y^+ = S_y^{e+} + S_y^{m+}; \tag{24}$$

$$S_z^{e+} = -S_z^{m+} = \frac{g}{\omega\mu}\sinh\alpha\,\mathrm{Re}\left(T_\| T_\perp^* E_{0\|} E_{0\perp}^*\right)e^{-2kz\sinh\alpha}, \quad S_z^+ = 0. \tag{25}$$

Remarkably, the electric and magnetic contributions to the $x$-component (23) are completely equivalent, whereas their counterparts in the $z$-oriented spin density exactly cancel each other producing the quite expected zero result. The contribution $S_x^+$ appears due to certain ellipticity of the evanescent wave polarization reflected by non-zero value of $\mathrm{Im}\left(T_\| T_\perp^* E_{0\|} E_{0\perp}^*\right) = \mathrm{Im}\left(E_{r\|}E_{r\perp}^*\right)$ and $S_z^{(e,m)+}$ owes to the mixture of $s$- and $p$-polarizations. It is Eq. (24) that describes the already mentioned transverse spin that is specific for an evanescent wave and constitutes a particular interest. In contrast to the usual propagating waves whose spin is collinear to the propagation direction, the component $S_y^+$ is directed orthogonally to the wavevector (9); moreover, it arises independently of the state of polarization of the incident wave. Now this feature, recently discovered in the surface polariton waves [18], is extended for a wide class of evanescent waves accompanying the total-reflection phenomena, and the study of its possible mechanical action will be considered in the subsequent sections.

In the medium 1 $(-\infty < z < 0)$ analogous calculations involving Eqs. (1), (2) and (14) yield

$$S_x^- = -2\frac{g}{\omega\mu_1}\sin\theta_1\,\mathrm{Im}\left[E_{0\perp}^* E_{0\|}\left(1+R_\perp^* e^{2ik_1 z\cos\theta_1}+R_\|e^{-2ik_1 z\cos\theta_1}+R_\perp^* R_\|\right)\right], \tag{26}$$

$$S_y^- = -2\frac{g}{\omega\mu_1}\sin\theta_1\cos\theta_1\,\mathrm{Im}\left[\left(\left|E_{0\|}\right|^2 R_\|^*+\left|E_{0\perp}\right|^2 R_\perp^*\right)e^{2ik_1 z\cos\theta_1}\right], \tag{27}$$

$$S_z^- = -2\frac{g}{\omega\mu_1}\cos\theta_1\,\mathrm{Im}\left[E_{0\perp}^* E_{0\|}\left(1-R_\|^* R_\|\right)\right]. \tag{28}$$

Here the electric and magnetic contributions are not explicitly separated because we do not intend to inspect interactions of this field with any material object, for which this separation can make sense. But confronting Eqs. (26) – (28) with the above Eqs. (23) – (25) may be interesting. First to note is that here, again, the non-zero transverse spin $S_y^-$ exists; here it looks not so bizarre as $S_y^+$ since the field in the medium 1 is a superposition of two plane waves where the transverse spin was described elsewhere [31,36]. However, its clear relation with the transverse spin of the evanescent wave (24) makes the nature and origination of the latter more understandable.

Second, all components of $\mathbf{S}$ depend only on $z$, so the general equation (15) can be reduced to the simple prescription

$$\mathbf{p}_S = \frac{1}{2}\left(-\mathbf{e}_x \frac{\partial S_y}{\partial z} + \mathbf{e}_y \frac{\partial S_x}{\partial z}\right). \tag{29}$$

Third and more important is that at the boundary ($z = 0$) quantities $S_{x,y,z}^+$ and $S_{x,y,z}^-$, in general, do not coincide. This peculiarity, also observed in the surface polariton field [20], means that, together with the "volume" spin momentum associated with quantities $S_{x,y}^+$ and $S_{x,y}^-$ separately, there exist the specific boundary contribution owing to the spin density discontinuity at $z = 0$,

$$\mathbf{p}_S^B = \mathbf{P}_S^B \delta(z), \quad \mathbf{P}_S^B = \frac{1}{2}\left[-\mathbf{e}_x\left(S_y^+ - S_y^-\right)_{z=0} + \mathbf{e}_y\left(S_x^+ - S_x^-\right)_{z=0}\right] \tag{30}$$

where $\delta(z)$ is the Dirac delta-function. These singular terms can hardly be observed but they are of principal value. In particular, they ensure that the total spin momentum of the field vanishes, which is required by the general spin momentum theory [20,21] (see Appendix).

Substituting Eqs. (23), (24) and (26) – (28) into Eq. (29) we obtain the following expressions for the "volume" spin momentum density contributions:

$$\begin{Bmatrix} p_{Sx}^{e+} \\ p_{Sx}^{m+} \end{Bmatrix} = -\frac{gn}{c\mu}\sinh^2\alpha\cosh\alpha \begin{Bmatrix} \left|T_\| E_{0\|}\right|^2 \\ \left|T_\perp E_{0\perp}\right|^2 \end{Bmatrix} e^{-2kz\sinh\alpha}, \quad p_{Sx}^+ = p_{Sx}^{e+} + p_{Sx}^{m+} \tag{31}$$

$$p_{Sy}^{e+} = p_{Sy}^{m+} = \frac{gn}{c\mu}\sinh\alpha\cosh\alpha\,\mathrm{Im}\left(T_\| T_\perp^* E_{0\|} E_{0\perp}^*\right)e^{-2kz\sinh\alpha}, \quad p_{Sy}^+ = 2p_{Sy}^{e+} = 2p_{Sy}^{m+} \tag{32}$$

$$p_{Sx}^- = 2\frac{gn_1}{c\mu_1}\sin\theta_1\cos^2\theta_1\,\mathrm{Re}\left[\left(\left|E_{0\|}\right|^2 R_\|^* + \left|E_{0\perp}\right|^2 R_\perp^*\right)e^{2ik_1 z\cos\theta_1}\right]. \tag{33}$$

$$p_{Sy}^- = -\frac{gn_1}{c\mu_1}\sin 2\theta_1\,\mathrm{Re}\left[E_{0\perp}^* E_{0\|}\left(R_\perp^* e^{2ik_1 z\cos\theta_1} - R_\|^* e^{-2ik_1 z\cos\theta_1}\right)\right], \tag{34}$$

Now, the orbital momentum components can be found directly via equation $\mathbf{p}_O = \mathbf{p} - \mathbf{p}_S$. Comparison of Eqs. (32) and (17), (34) and (19) shows that $p_{Sy}^+ = p_y^+$, $p_{Sy}^- = p_y^-$ and, consequently, $p_{Oy}^+ = p_{Oy}^- = 0$: the "volume" transverse energy flow is completely of the spin nature. The $z$-component of the spin momentum vanishes due to Eq. (29), which in combination with Eq. (20) means that $p_{Oz} = 0$ in the entire space. The whole "volume" orbital momentum is thus $x$-oriented and, according to Eqs. (18), (16) and (31), (33), equals to

$$p_{Ox}^+ = \frac{gn}{c\mu}\cosh^3\alpha\left(\left|T_\perp E_{0\perp}\right|^2 + \left|T_\| E_{0\|}\right|^2\right)e^{-2kz\sinh\alpha}, \tag{35}$$

$$p_{Ox}^- = 2\frac{gn_1}{c\mu_1}\sin\theta_1\left\{\left|E_{0\|}\right|^2 + \left|E_{0\perp}\right|^2 + \sin^2\theta_1\,\mathrm{Re}\left[\left(\left|E_{0\|}\right|^2 R_\|^* + \left|E_{0\perp}\right|^2 R_\perp^*\right)e^{2ik_1 z\cos\theta_1}\right]\right\}. \tag{36}$$

Note that the $x$-directed spin and orbital momentum components in the evanescent-wave region are directed oppositely and obey the relations

$$p_{Sx}^+ = -\sinh^2\alpha \cdot p_x^+, \quad p_{Ox}^+ = \cosh^2\alpha \cdot p_x^+,$$

similar to what was found for the surface polariton [20] and interpreted by means of the idea that the "negative" spin momentum ensures proper (subluminal) energy transport in the evanescent field. The spin part of the $x$-directed momentum (31) and (33) is associated with the transverse spin density $S_y^+$ (24) and $S_y^-$ (27). It arises due to the field inhomogeneity and elliptic polarization in plane ($XZ$), which may appear in the reflected or refracted field even if the incident wave is linearly polarized in $s$- or $p$-direction.

Eqs. (31) – (36) do not provide a complete description of the field momentum. The boundary contributions defined in Eqs. (30) should be added whose expressions follow directly from Eqs. (23), (24) and (26), (27):

$$P_{Sy}^B = -\frac{g}{\omega\mu}\cosh\alpha\,\mathrm{Im}\left(T_{\parallel}T_{\perp}^* E_{0\parallel}E_{0\perp}^*\right) + \frac{g}{\omega\mu_1}\sin\theta_1\,\mathrm{Im}\left[E_{0\perp}^* E_{0\parallel}\left(1 + R_{\perp}^* + R_{\parallel} + R_{\perp}^* R_{\parallel}\right)\right]$$

$$= -\frac{g}{\omega\mu}\left(\frac{n_1}{n} - \frac{n}{n_1}\right)\sin\theta_1\,\mathrm{Im}\left(T_{\perp}^* T_{\parallel} E_{0\perp}^* E_{0\parallel}\right); \tag{37}$$

$$P_{Sx}^B = \frac{g}{2\omega\mu}\sinh\alpha\cosh\alpha\left(\left|T_{\perp}E_{0\perp}\right|^2 + \left|T_{\parallel}E_{0\parallel}\right|^2\right) - \frac{g}{\omega\mu_1}\sin\theta_1\cos\theta_1\,\mathrm{Im}\left[\left(\left|E_{0\parallel}\right|^2 R_{\parallel}^* + \left|E_{0\perp}\right|^2 R_{\perp}^*\right)\right]$$

$$= -\frac{g}{\omega}\frac{\varepsilon_1}{n n_1}\left(\frac{n_1}{n} - \frac{n}{n_1}\right)\cos\theta_1\sin\theta_1\left(\left|E_{0\parallel}\right|^2\mathrm{Im}\,R_{\parallel} + \left|E_{0\perp}\right|^2\mathrm{Im}\,R_{\perp}\right). \tag{38}$$

In the latter transformations, we have employed Eqs. (5) and relations

$$1 + R_{\perp} = T_{\perp}, \quad 1 + R_{\parallel} = \sqrt{\frac{\varepsilon\mu_1}{\varepsilon_1\mu}}T_{\parallel}, \quad \left|T_{\parallel}\right|^2 = 2\frac{\cos\theta_1}{\sinh\alpha}\sqrt{\frac{\varepsilon_1\mu}{\varepsilon\mu_1}}\mathrm{Im}\,R_{\parallel}^*, \quad \left|T_{\perp}\right|^2 = 2\frac{\cos\theta_1}{\sinh\alpha}\sqrt{\frac{\varepsilon_1\mu}{\varepsilon\mu_1}}\mathrm{Im}\,R_{\perp}^*$$

between the reflection (4) and transmission (8) coefficients.

Since the total electromagnetic momentum (16) – (20) contains no singularities, the spin momentum singular terms (37) and (38) imply that the similar, but oppositely directed, singular boundary flows of the orbital nature also exist. This means that, in contrast to the linearly polarized field of the surface polariton [20], generally one cannot assert that the whole field momentum orthogonal to the plane of incidence is of the spin nature: the specific boundary spin flow (37) is accompanied by the oppositely directed "compensatory" orbital flow. The form of Eqs. (32), (34) and (37) dictates that any $y$-directed field momentum, no matter of the spin or orbital nature, appears due to elliptic polarization of at least one of interacting waves: incident, reflected or refracted (evanescent).

### 3. Calculation of the field mechanical action

Following to the procedure of Ref. [34], we can numerically calculate the scattered field $\mathbf{E}^s$, $\mathbf{H}^s$ arising due to interaction between the incident evanescent wave in the form (6) and a spherical particle of radius $a$ with electromagnetic parameters $\varepsilon_p$, $\mu_p$, and $n_p = \sqrt{\varepsilon_p\mu_p}$. The procedure involves only the standard Mie theory equations [37] and establishes linear relation between the scattered field and the evanescent wave amplitude, which can be written in a symbolic operator form as

$$\mathbf{E}^s(\mathbf{r}) = \hat{S}_E(\mathbf{r})\mathbf{E}_t, \quad \mathbf{H}^s(\mathbf{r}) = \hat{S}_H(\mathbf{r})\mathbf{E}_t, \tag{39}$$

where matrices $\hat{S}_{E,H}(\mathbf{r})$ are composed from the Mie scattering operators and complex-angle spatial rotation operators [34]. Eqs. (39) imply that the frame origin coincides with the particle center; if the latter is removed from the interface by $h$, $\mathbf{E}_t$ is determined by Eq. (6) with $z = h$. The total field is then given by the vector summation of the incident and scattered fields:

$$\mathbf{E}^{tot} = \mathbf{E}_t + \mathbf{E}^s, \quad \mathbf{H}^{tot} = \mathbf{H}_t + \mathbf{H}^s. \tag{40}$$

Once the total field is known, the mechanical action can be calculated via the standard procedures based on the Maxwell stress tensor $\hat{T} = \{T_{ij}\}$, $i, j = x, y, z$:

$$T_{ij} = g \, \mathrm{Re}\left[ \varepsilon E_i^{tot*} E_j^{tot} + \mu H_i^{tot*} H_j^{tot} - \frac{1}{2}\delta_{ij}\left( \varepsilon \left|\mathbf{E}^{tot}\right|^2 + \mu \left|\mathbf{H}^{tot}\right|^2 \right) \right] \tag{41}$$

and the field angular momentum flux tensor [38] $\hat{M} = \{M_{li}\}$, $M_{li} = g \epsilon_{ijk} x_j T_{kl}$ where $\epsilon_{ijk}$ is the Levi-Civita symbol. Integrating the stress tensor and the angular momentum flux tensor components over any surface $A$ enclosing the particle (e.g., a sphere $S = \{r = R\}$, $R > a$), we obtain the optical force and torque:

$$\mathbf{F} = \oint_A \hat{T}\,\mathbf{n}\,dA = R^2 \int_S \hat{T}\,\mathbf{n}\,d\Omega, \quad \mathbf{Q} = \oint_A \hat{M}\,\mathbf{n}\,dA = R^2 \int_S \hat{M}\,\mathbf{n}\,d\Omega \tag{42}$$

where $d\Omega = \sin\theta\, d\theta\, d\phi$ is the elementary solid angle, $\mathbf{n} = (\sin\theta\cos\phi, \sin\theta\sin\phi, \cos\theta)^T$ is the unit vector of the outer normal to the sphere surface.

For the numerical evaluations we assume that a spherical particle of radius $a$ lies on the totally-reflecting surface ($h = a$). Other necessary input parameters are assumed as in [34]:

$$\mu_1 = \mu = \mu_p = 1, \quad n_1 = 1.75, \quad \theta_1 = 51°. \tag{43}$$

To reduce the dynamical range of the presented data and realizing the conditions of the earlier simulations [34], the calculated force and torque values are normalized via dividing by the quantities

$$P_0 = \frac{a^2}{4\pi}\left( \left|E_{0\perp}\right|^2 + \left|E_{0\|}\right|^2 \right), \quad Q_0 = \frac{P_0}{k}, \tag{44}$$

respectively, which are proportional to the time-average momentum flux through the area $a^2$ within the field of the plane wave (1) approaching the interface. We consider two variants of the evanescent-region media: water

$$n = 1.33, \quad \alpha = 0.212, \quad \theta_c = 49.46°, \quad T_\| = 2.19 - 0.98\mathrm{i}, \quad T_\perp = 1.88 - 0.48\mathrm{i} \tag{45}$$

and air/vacuum

$$n = 1.0, \quad \alpha = 0.825, \quad \theta_c = 34.85°, \quad T_\| = 0.46 - 1.19\mathrm{i}, \quad T_\perp = 1.18 - 0.98\mathrm{i} \tag{46}$$

($\theta_c$ denotes the total-reflection critical angle). Most of the results are obtained under conditions when the incident wave in the high-index region $z < 0$ is circularly polarized

$$E_{0\perp} = i\sigma E_{0\|}, \quad \sigma = \pm 1. \tag{47}$$

In all figures (see below) the results obtained for $\sigma = +1$ (−1) are presented by solid (dashed) curves.

## 3.1. Optical forces in the evanescent field

Figs. 1 and 2 show results of the force calculation for particles with different electric properties and variable radius $a$ expressed via parameter $ka$. The forces parallel to the plane of incidence (vertical $F_z$ and horizontal $F_x$) behave quite expectedly, showing the features known from the calculations for *s*- and *p*-polarizations of the incident wave (1) performed earlier for the same conditions (43) – (46) and the same particles [34]. Even in conditions of Fig. 2a where, in order to characterize the particle absorptive properties, we replaced the zero imaginary part of the perfect-metal permittivity by a small positive value, this does not entail significant modifications of curves $F_z$ and $F_x$ against their counterparts in [34] (except some quantitative distinctions caused by the different incident polarization). In Figs. 1b and 2b, due to the higher difference in refraction indices of the particle and the medium, all forces become stronger and their dependence on the model parameters acquires the oscillatory character which is explained by the morphology-dependent resonances in the particle with $ka > 1$ [37].

The most impressive novelty is the rather strong transverse force $F_y$ whose absolute value may reach tens of percents of the traditional longitudinal force $F_x$ (remarkably, this roughly corresponds to the usual proportions between the Imbert–Fedorov and Goos–Hänchen beam shifts [15–17]). Contrary to the other components, $F_y$ changes the sign upon switching the polarization handedness (this is shown explicitly in Fig. 1a; in Figs. 1b as well Fig. 2a, b only the curves for one handedness, more suitable for the demonstration, are presented). This allows one to consider the force $F_y$ as a mechanical manifestation of the transverse energy flow (32) that is of the spin nature. In view of Eq. (47), this spin momentum can be represented as

$$p_{Sy}^{e+} = p_{Sy}^{m+} = -\frac{gn}{c\mu}\sigma\left|E_{0\parallel}\right|^2 \sinh\alpha\cosh\alpha\,\mathrm{Re}\left(T_{\parallel}T_{\perp}^*\right)e^{-2kz\sinh\alpha}. \qquad (48)$$

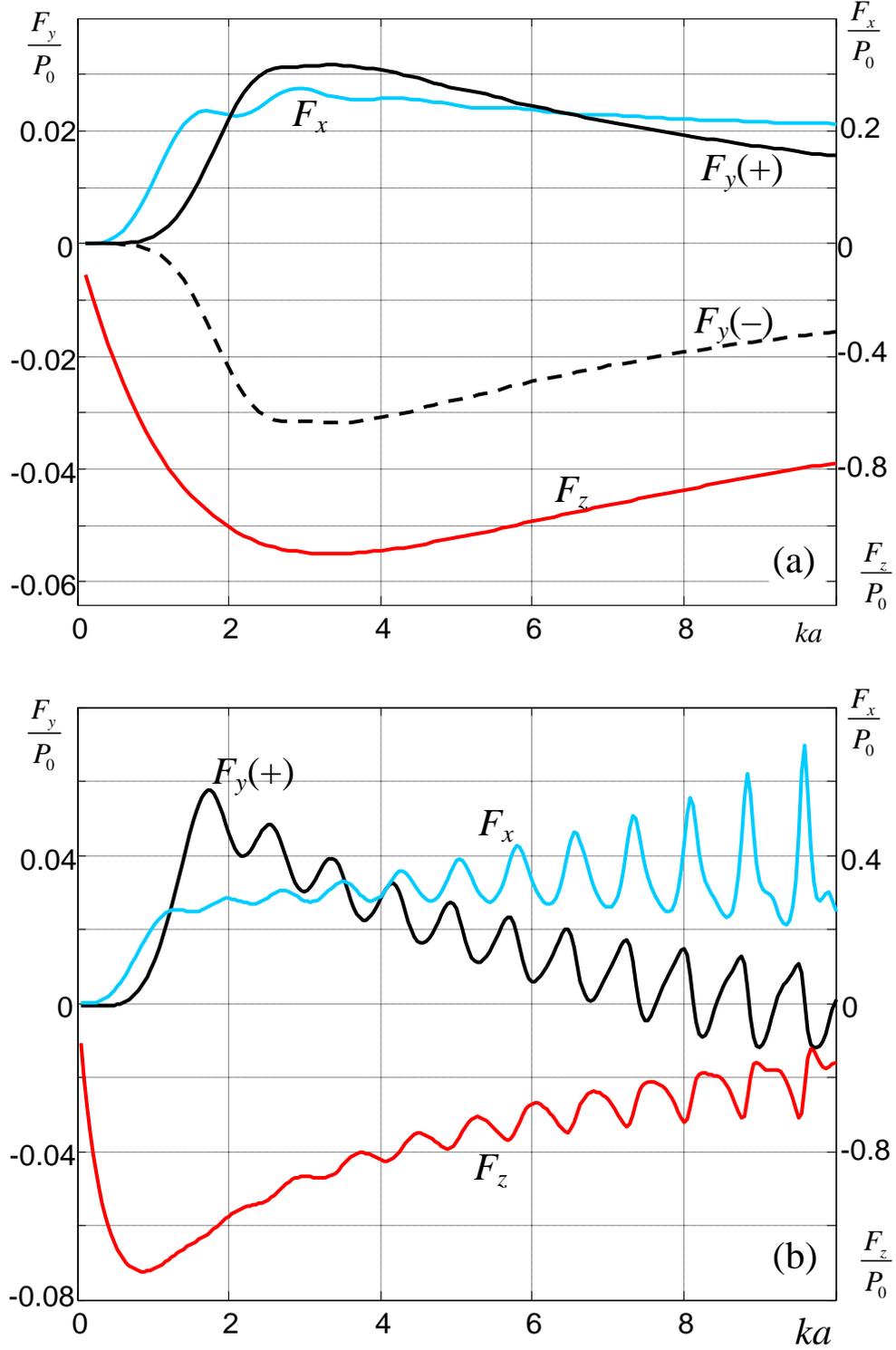

Fig. 1. Normalized radiation force components vs the particle size parameter, calculated for dielectric particle with $n_p = 1.5$ under conditions (43) in case of circular polarization of Eq. (47): (a) $n = 1.33$, $\sigma = \pm 1$; (b) $n = 1$, $\sigma = 1$. Forces $F_x$ and $F_z$ (right vertical scales) do not depend on the polarization handedness, $F_y$ (left vertical scale) changes the sign with $\sigma$ (this is shown by the solid $F_y(+)$ and dashed $F_y(-)$ black curves in panel (a); in panel (b) only the case of positive helicity $\sigma = 1$ [curve $F_y(+)$] is shown).

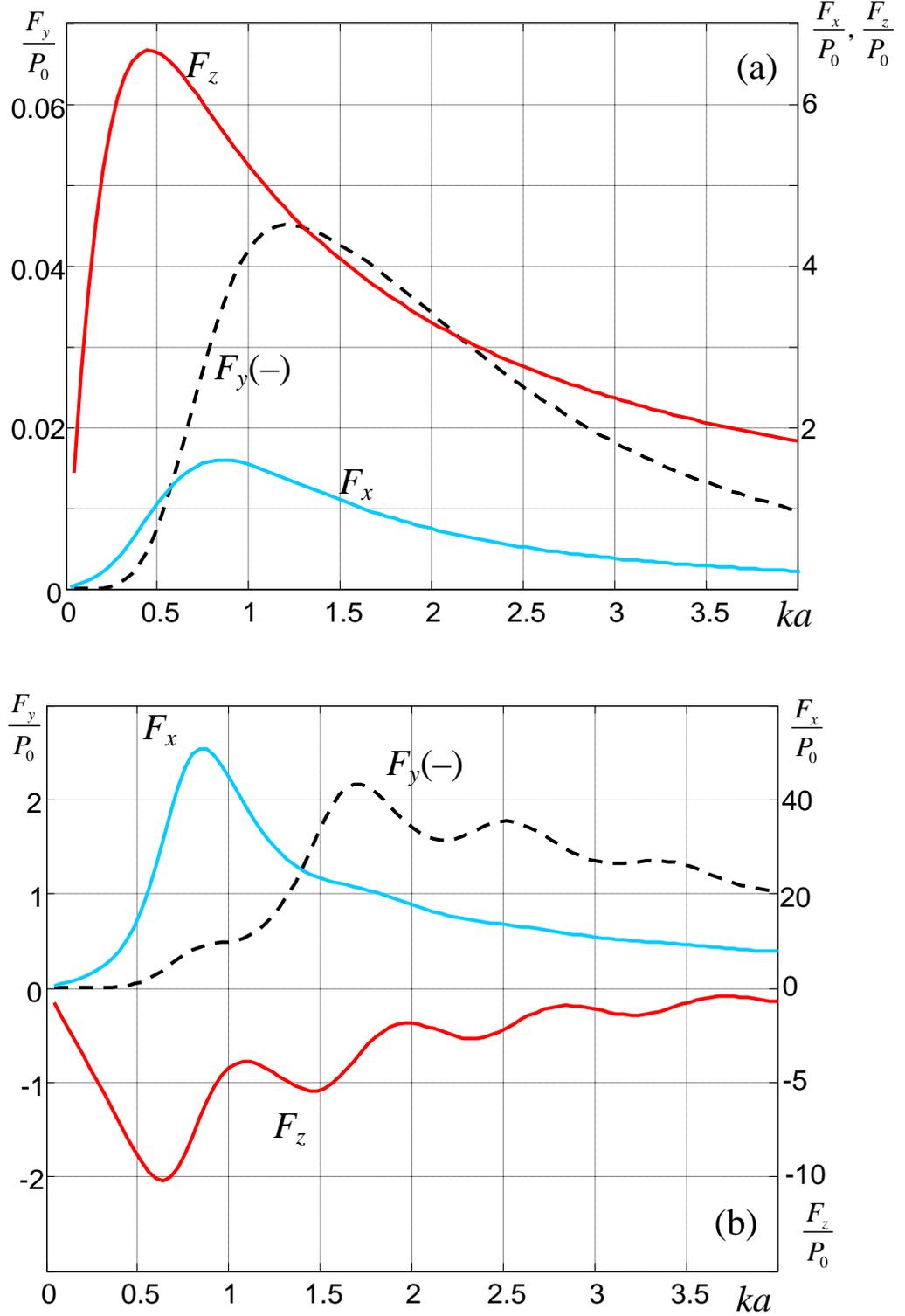

Fig. 2. Normalized radiation force components vs the particle size parameter $ka$, calculated for conditions (43) in case of circular polarization of Eq. (47), $\sigma = -1$: (a) $n = 1$, $\varepsilon_p = -1 + 0.01\text{i}$ (ideal conducting particle with small absorption); (b) $n = 1.33$, $n_p = 0.43 + 3.52\text{i}$ (gold particle in water at $\lambda = 650$ nm [39]). For $\sigma = +1$, curves $F_y$ (left vertical scales) are reflected with respect to the horizontal axis, other curves (right vertical scales) do not change.

Under conditions (45) and (46), $\mathrm{Re}\left(T_{\parallel}T_{\perp}^*\right) > 0$, and the sign of the whole expression (48) is opposite to $\sigma$. That is, in contrast to the metallic particle behavior in Fig. 2, on a small dielectric particle (at least, while $ka < 6$ in Fig. 1) this force acts against its "source" – the spin momentum (48), which is also a characteristic feature of the spin-momentum induced mechanical action [31–33].

As expected for the spin-momentum force exerted on a non-magnetic particle [33], at $ka \ll 1$ curves $F_y$ grow as $a^8$. In contrast, the horizontal force $F_x$ scales up with $a^6$ in both panels of Fig. 1, which is typical for the orbital flow action on particles with the real polarizability [30,33]; in Fig. 2 in the small-$ka$ region $F_x$ grows proportionally to $a^3$ due to the non-zero imaginary part of the complex permittivity $\varepsilon_p$. The direction of $F_x$ always agrees with the direction of the orbital momentum density $p_{Ox}$ (35). The spin momentum (31) might also contribute to $F_x$ but its effect does not depend on $\sigma$ and, generally, cannot be separated from the orbital momentum contribution within the frame of the present numerical analysis. However, from the general notions on the spin-momentum ponderomotive action [33], one might expect the spin contribution to $F_x$ to be negligibly small for the Raylegh-scattering particles with $ka \ll 1$. The vertical force $F_z$ shows the characteristic gradient-force behavior (~$a^3$ in the low-size region) [33], which is not surprising since the $z$-directed momentum of the analyzed field is zero (see Eq. (20)) and the total $z$-directed action is completely due to inhomogeneous intensity.

## 3.2. Optical torques in the evanescent field

In this section we inspect mechanical consequences of the unusual transverse spin (24) carried by the evanescent field. Our calculations predictably have shown that non-absorbing particles do not feel the field spin, so Figs. 3 and 4 only present the results obtained for the conductive particles previously considered in Fig. 2.

First and quite expected detail of both figures is the presence of the longitudinal torque $Q_x$ stipulated by circular polarization of the incident wave and changing the sign with the polarization handedness (blue curves in panels 3a and 4a), which can be related to the spin component $S_x$ (23). Also, graphs in Figs. 3 and 4 confirm the existence of a $y$-directed torque associated with the transverse spin (24) (curves $Q_y$) and vanishing of all other torque components in cases of "pure" $s$- or $p$-polarization, in full compliance with Eqs. (23) and (25). In agreement with Eqs. (24), the $Q_y$ component does not depend on the polarization handedness. However, it appears to be much more sensitive to the $p$-polarized incident wave (1) than to the $s$-polarized one (even the torque $Q_y(+,-)$ in Figs. 3a and 4a is approximately two times lower than the torque $Q_y(p)$ in Figs. 3b, 4b, which closely corresponds to the relative weight of the $p$-polarization in the incident circularly polarized radiation). Still more striking, in view of the vanishing $z$-component of the field spin (25), looks the non-zero $z$-oriented torque [curves $Q_z(+)$, $Q_z(-)$] whose direction switches together with the polarization handedness.

Both the essential difference between curves $Q_y(p)$ and $Q_y(s)$ and the quite perceptible "unexpected" $Q_z(+)$, $Q_z(-)$ components apparently cannot be linked to the "whole" spin components described by the second Eqs. (24), (25). However, these can be fairly explained by the selective sensitivity of the probe particles to the electric and magnetic contributions exposed by the first Eqs. (24), (25). In fact, both sorts of particles satisfying first condition (43) are preferentially subject to electric, rather than magnetic, interactions. That is why a particle does not "feel" the total spin $S_z^+ = 0$ but "extracts" its electric part that in view of Eq. (47) behaves as

$$S_z^{e+} = \frac{g}{\omega\mu}\sigma\left|E_{0\parallel}\right|^2 \sinh\alpha\, \mathrm{Im}\left(T_\parallel T_\perp^*\right)e^{-2kz\sinh\alpha} \tag{49}$$

and since $\mathrm{Im}\left(T_\parallel T_\perp^*\right) > 0$ [cf. Eqs. (45) and (46)], its sign coincides with the sign of $\sigma$ (which can be clearly observed in curves $Q_z(+)$, $Q_z(-)$ of Figs. 3a and 4a). Likewise, the electric part of the transverse spin (25) only depends on the $p$-polarized incident wave component and is therefore almost insensitive to the orthogonal $s$-polarized light (cf. curves $Q_y(p)$ and $Q_y(s)$ in Figs. 3b and 4b).

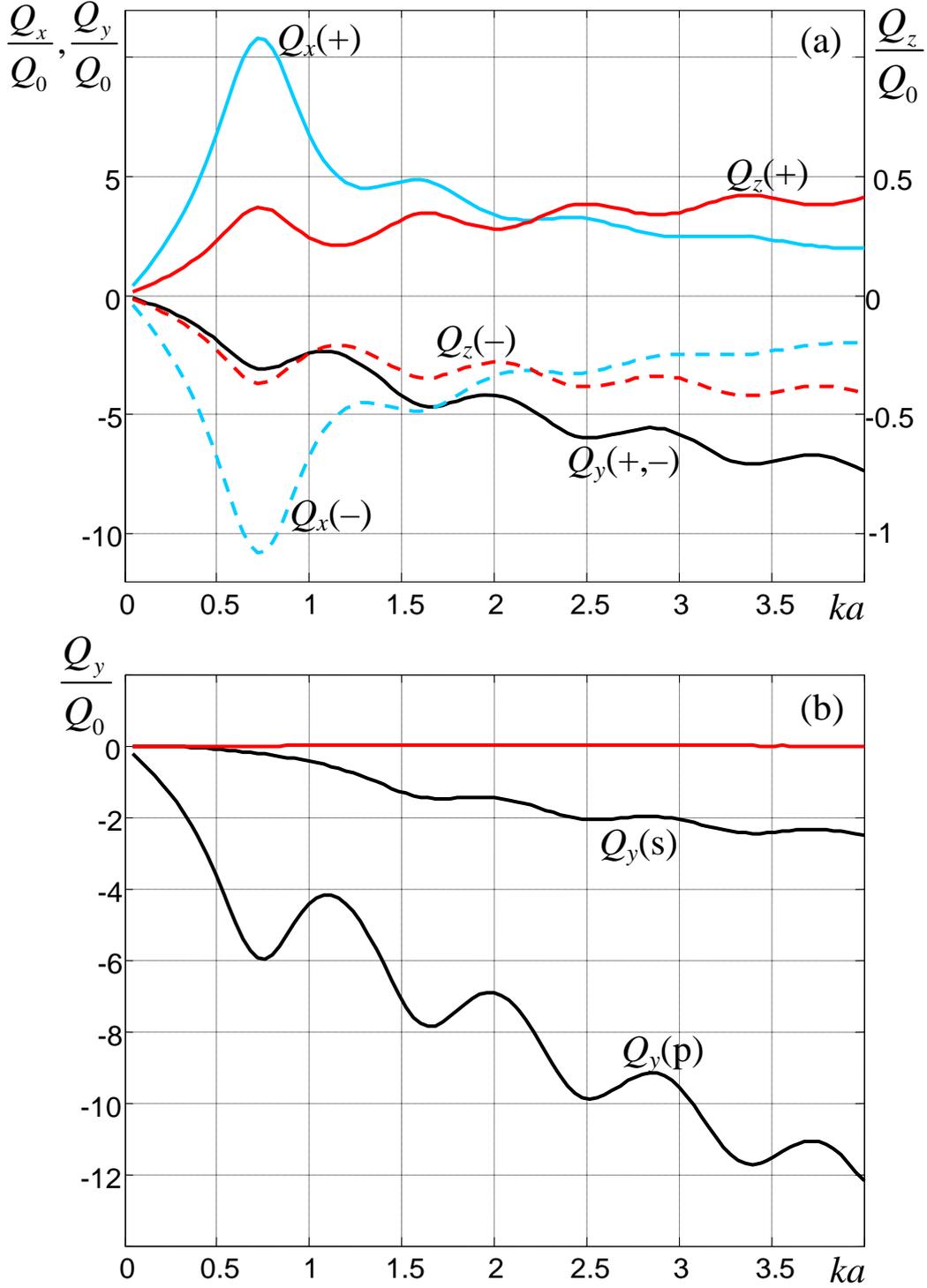

Fig. 3. Normalized radiation torque components vs the particle size parameter *ka*, calculated for conditions (43) and *n* = 1.33, $n_p$ = 0.43 + 3.52i (gold particle in water at $\lambda$ = 650 nm [39]): (a) circular polarization of Eq. (47), handedness is marked by "+" or "−"; (b) two cases of linear polarization (marked "s" and "p"). In panel (a) torques $Q_x$ and $Q_z$ change the sign with switching the polarization handedness whereas $Q_x$ remains the same; in panel (b) all curves for $Q_x$ and $Q_z$ visually coincide with the horizontal axis.

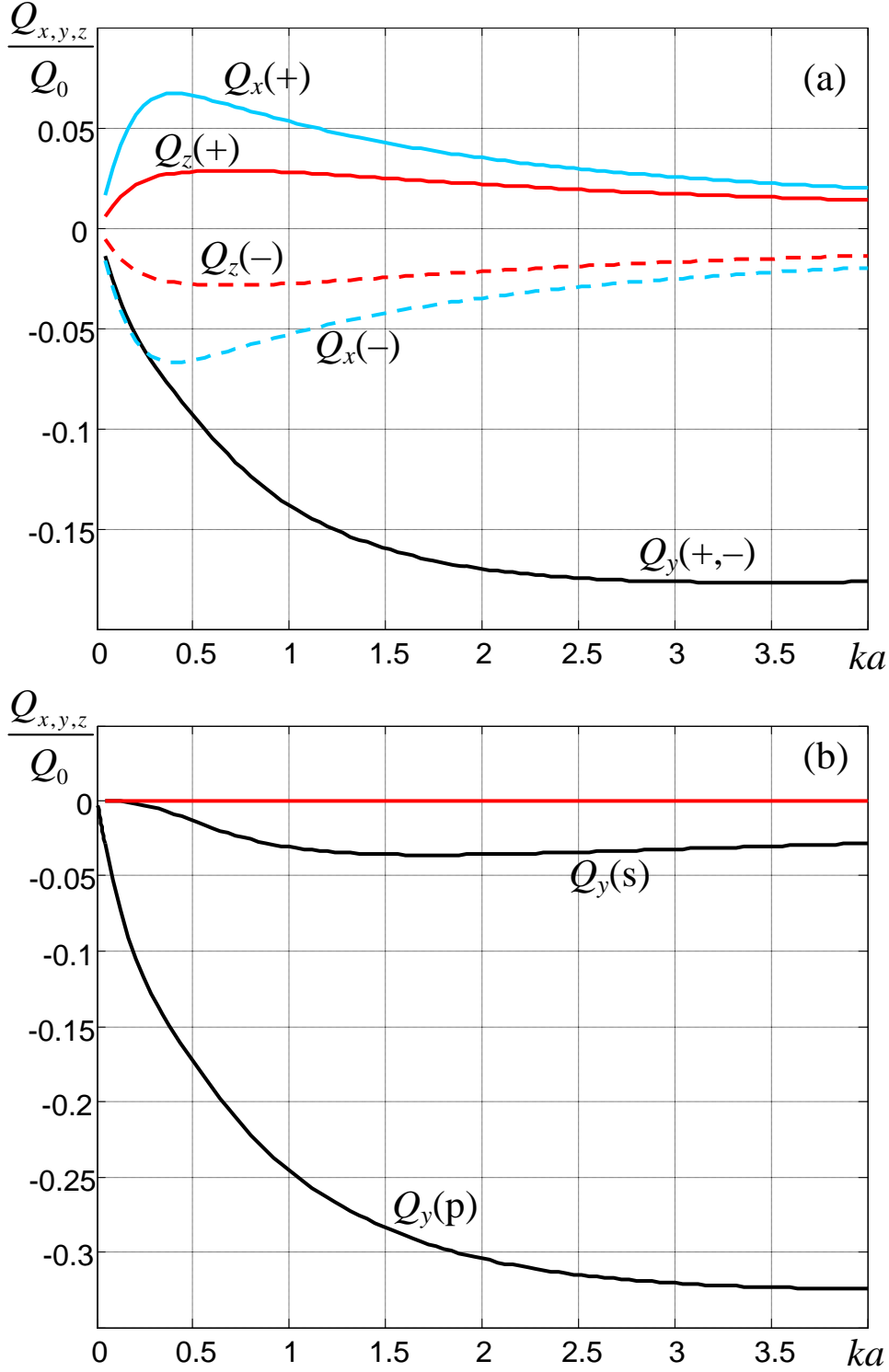

Fig. 4. Normalized radiation torque components vs the particle size parameter $ka$, calculated for conditions (43) and $n = 1$, $\varepsilon_p = -1 + 0.01\mathrm{i}$ ("ideal" metal with small absorption): (a) circular polarization of Eq. (47), handedness is marked by "+" or "–"; (b) two cases of linear polarization (marked "s" and "p"). In panel (a) torques $Q_x$ and $Q_z$ change the sign with switching the polarization handedness whereas $Q_x$ remains the same; in panel (b) all curves for $Q_x$ and $Q_z$ visually coincide with the horizontal axis.

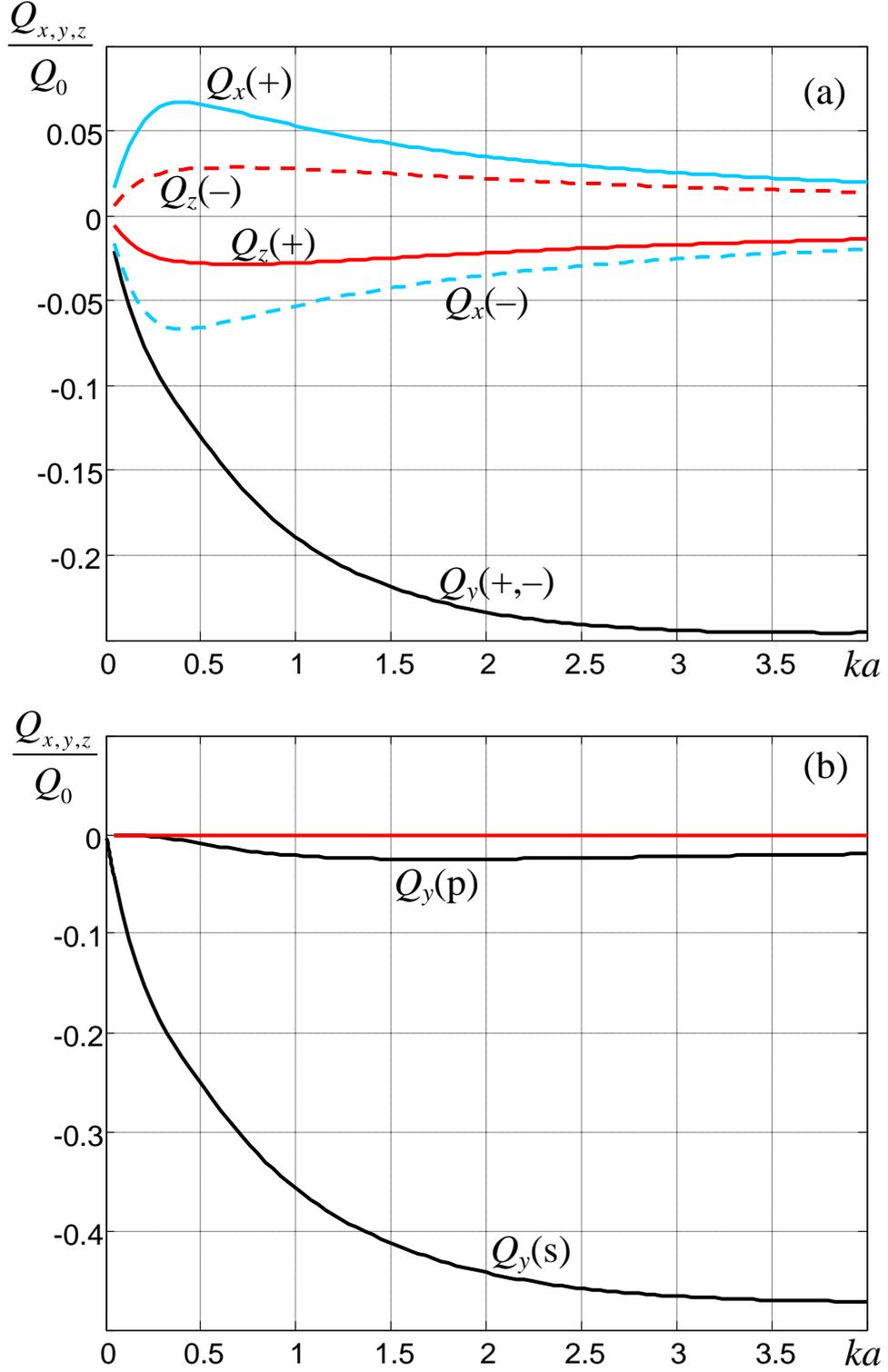

Fig. 5. Same as Fig. 4 but for magnetic particle with $\varepsilon_p = 1$, $\mu_p = -1 + 0.01\mathrm{i}$. Compared to Fig. 4, in panel (a) the torque components $Q_z$ caused by circular polarization (47) change the sign, and in case of linear polarization (b) the torque $Q_y$ is more sensitive to the $s$-polarized incident wave than to the $p$-polarized one.

These considerations are confirmed by calculations performed at the same conditions (43) and (45), (46) but for the "dual" particle with $\varepsilon_p = 1$, $\mu_p = -1 + 0.01\mathrm{i}$ (Fig. 5). In this case, on the contrary, the magnetic parts of the spin components (23) – (25) act as "motive agents". This expectedly makes little difference for $Q_x(+)$ and $Q_x(-)$ as the electric and magnetic parts of the $x$-directed spin (23) are equal but for other torque contributions the situation looks almost mirror-like: signs of $Q_z(+)$ and $Q_z(-)$ in Fig. 5a are now opposite to $\sigma$ [obvious consequence of the electric – magnetic "opposition" in the first Eq. (25)] and the torque $Q_y(s)$ essentially exceeds $Q_y(p)$ (compare Figs. 4b and 5b). The incomplete numerical correspondence is explained by "incomplete duality" of the situations of Figs. 4 and 5: to make it accomplished, one should also interchange the electric and magnetic constants of the involved medium 1).

Following to the known approach [33], employed already in the end of Sec. 3.1, some conclusions on the physical origins of the field-induced torques of Figs. 3 – 5 can be inferred from their behavior in the $ka \ll 1$ limit. All the quantities presented in these figures, except $Q_y(s)$ in Fig. 4b and $Q_y(p)$ in Fig. 5b, grow as $\sim a^3$; besides, they turn out to be proportional to the absorption index. This means that the particles, absorbing a part of the light energy, acquire also the torque carried by this light, each component with its own "sensitivity" depending on the particle electric or magnetic properties, but always proportionally to the energy absorbed. All these effects are described by the leading terms of the Mie series [37] expressing the scattered field as the field of an induced dipole (electric or magnetic depending on the electric or magnetic nature of the particle). The "odd" torques $Q_y(s)$ in Figs. 3b and 4b and $Q_y(p)$ in Fig. 5b make an exclusion. At small $ka$, they scale up with $a^5$, which means that they appear in higher orders of the multipole expansion. The matter is that, say, $s$-polarized wave with the spin density (24) cannot induce the electric dipole moment in a non-magnetic particle but induces the much weaker electric quadrupole and magnetic dipole moments (described by higher-order terms of the Mie series) that eventually are responsible for the torque imparted to the particle.

### Appendix

The total spin momentum of the field consists of the $x$- and $y$-components which are proportional to the following integrals taken over the whole $z$-axis:

$$P_{Sx,Sy}^{tot} = \int_{-\infty}^{\infty} p_{Sx,Sy}(z)\,dz = \int_{-\infty}^{0} p_{Sx,Sy}^-(z)\,dz + \int_{0}^{\infty} p_{Sx,Sy}^+(z)\,dz + P_{Sx,Sy}^B \tag{A1}$$

where the last term is obtained via integration of the delta-function in (30). The direct evaluation of expression (A1) is impossible because the first integral in the right-hand side with $p_{Sx,Sy}^-(z)$ taken in form (33), (34) diverges. To make the integral meaningful, the integrand quantities $p_{Sx,Sy}^-(z)$ should be properly regularized. To this end, we imply the infinitesimal absorption in the medium 1 so that Eqs. (26) and (27) are replaced by

$$S_x^- = 2\frac{g}{\omega\mu_1}\sin\theta_1\, \mathrm{Im}\left[E_{0\perp}^* E_{0\parallel}\left(1 + R_\perp^* e^{2ik_1 z\cos\theta_1} + R_\parallel e^{-2ik_1 z\cos\theta_1} + R_\perp^* R_\parallel\right)\right]e^{\kappa z},\; \kappa \to 0. \tag{A2}$$

$$S_y^- = -2\frac{g}{\omega\mu_1}\sin\theta_1\cos\theta_1\,\mathrm{Im}\left[\left(\left|E_{0\|}\right|^2 R_\|^* + \left|E_{0\perp}\right|^2 R_\perp^*\right)e^{2ik_1 z\cos\theta_1}\right]e^{\kappa z},\ \kappa\to 0; \tag{A3}$$

Substituting this into Eq. (29) one finds, instead of Eqs. (33) and (34),

$$p_{Sy}^- = \frac{gn_1}{c\mu_1}\sin\theta_1\,\mathrm{Im}\left\{E_{0\perp}^* E_{0\|}\left[\frac{\kappa}{k_1}\left(1 + R_\perp^* R_\|\right) + \left(2i\cos\theta_1 + \frac{\kappa}{k_1}\right)R_\perp^* e^{2ik_1 z\cos\theta_1}\right.\right.$$
$$\left.\left. - \left(2i\cos\theta_1 - \frac{\kappa}{k_1}\right)R_\| e^{-2ik_1 z\cos\theta_1}\right]\right\}e^{\kappa z},\ \kappa\to 0; \tag{A4}$$

$$p_{Sx}^- = \frac{gn_1}{c\mu_1}\sin\theta_1\cos\theta_1\,\mathrm{Im}\left[\left(2i\cos\theta_1 + \frac{\kappa}{k_1}\right)\left(\left|E_{0\|}\right|^2 R_\|^* + \left|E_{0\perp}\right|^2 R_\perp^*\right)e^{2ik_1 z\cos\theta_1}\right]e^{\kappa z},\ \kappa\to 0. \tag{A5}$$

Now the integral over the $-\infty < z < 0$ half-space can be easily evaluated; other integral terms in the right-hand side of (A1) cause no problems because the decaying exponential factor contained in $p_{Sx,Sy}^+$ (31), (32) ensures their convergence. After the integrations are performed, a passage to the limit $\kappa\to 0$ and allowance for Eqs. (37) and (38) immediately entail that quantity (A1) vanishes and, therefore the total spin momentum of the field is zero.